\documentclass[twocolumn,showpacs,fleqn,nobibnotes]{revtex4}
\usepackage{amsmath}
\usepackage{graphicx}
\usepackage{float}
\usepackage{subfigure}
\usepackage{multirow}
\usepackage{color}

\def\lsim{\raise0.3ex\hbox{$<$\kern-0.75em\raise-1.1ex\hbox{$\sim$}}}
\def\gsim{\raise0.3ex\hbox{$>$\kern-0.75em\raise-1.1ex\hbox{$\sim$}}}

\begin{document}

\title{$J/\Psi$ PHOTOPRODUCTION IN PERIPHERAL AA COLLISIONS}
\author{M. B. GAY DUCATI\footnote{beatriz.gay@ufrgs.br} AND S. MARTINS\footnote{sony.martins@ufrgs.br}}

\affiliation{High Energy Physics Phenomenology Group, GFPAE  IF-UFRGS \\
Caixa Postal 15051, CEP 91501-970, Porto Alegre, RS, Brazil}

\begin{abstract}
The exclusive photoproduction of the heavy vector mesons $J/\Psi$ is investigated in the context of peripheral lead-lead collisions for the energies available at the LHC, $\sqrt{s}=2.76$ TeV and $\sqrt{s}=5.02$ TeV. Using the light-cone color dipole formalism, it was calculated the rapidity distribution in two centrality bins 50\%-70\% and 70\%-90\% in order to evaluate its robustness in extrapolating down to smaller impact parameter. It is introduced a modified photon flux, without change in the photonuclear cross section in relation to the ultraperipheral (UPC) case. Results were obtained for the two regions analyzed, which presented a maximum difference of 27\% in frontal rapidity for the two regions. Comparing the results for $\sqrt{s}=2.76$ TeV and $\sqrt{s}=5.02$ TeV, it was verified an increase of approximately half the one obtained in the ultraperipheral regime in the central rapidity region. 

\end{abstract}

\maketitle

\section{Introduction}

For a long time, the production of charmonium has been considered as a clean probe for the study of matter formed in high energy nuclear collisions \cite{PLB178-416}. In this limit, where the production of charm quarks is numerous and it is believed to occur the formation of the Quark-Gluon Plasma (QGP), two relevant effects are present: charmonium suppression and $c\bar{c}$ recombination. The first is associated with the nuclear medium temperature which becomes greater than the dissociation temperature of the charmonium, causing its destruction \cite{NPA610-404}. The second, also called regeneration, is characterized by the recombination process of initially uncorrelated charm quarks $c$ and $\bar{c}$ into a charmonium \cite{NPA709-415}. The consideration of these two effects is necessary for the understanding of the charmonium production at the Relativistic Heavy Ion Collider (RHIC) \cite{PRL97-232301,PLB664-253}. On the other hand, in the Large Hadron Collider (LHC), where the energy reaches an order of magnitude higher than in RHIC, the collisions with much smaller $x$ are in the strong shadowing region \cite{EPJC9-61,PLB226-167}, where the so called cold nuclear matter effects, as shadowing \cite{NPB268-427,NPB493-305,NPB511-355}, can significantly affect the charmonium production. Thus, in order to understand the charmonium production and extract properties of the medium created in high energy nuclear collisions, one must take into account both cold and hot nuclear matter effects.

The main way used to analyze all these effects is the calculation of the nuclear modification factor $R_{AA}$, which compares the final yield of charmonium from heavy ion collisions to that from the corresponding nucleon-nucleon collisions. In the last years, it has increased the interest in calculating the $R_{AA}$ as a function of multiplicity, transverse momentum and rapidity of the $J/\Psi$'s \cite{talk}. The ALICE collaboration, by measuring this observable as a function of transverse momentum, has pointed out an increase in the inclusive production of the $J/\psi$, at small $p_{T}$ ($p_{T}<300$ MeV/c), in the frontal rapidity region \cite{PRL116-222301}. One of the first hypotheses is that this excess could be originated from coherent photoproduction of the meson in the peripheral region \cite{PRL116-222301}. The photoproduction of heavy vector mesons has already been well explored in ultraperipheral collisions \cite{PRD94-094023,PRD88-017504,PRC89-025201,JPG42-105001,PRC87-032201,PRC93-055206,AIP1654,TMP182-141} and can act as a complement to allow us to obtain information about the gluon distribution in the nuclear medium. However, there are few works in the peripheral collisions regime (in particular, $b \simeq 2R_A$), where the exclusive photoproduction mechanism may still be relevant for the heavy vector mesons production. In \cite{PRC93-044912}, for example, this issue is addressed from a modification in the equivalent photon flux, without change in the photonuclear cross section in relation to the ultraperipheral case. Following this same idea, we tested the formalism used in our previous work \cite{PRD94-094023} by calculating the rapidity distribution for the coherent photoproduction of the $J/\Psi$ in Pb-Pb collisions in the centrality classes: 50\%-70\% and 70\%-90\%. Based in the good results obtained in the ultraperipheral regime, it was considered here the light-cone color dipole formalism \cite{nik}, which includes consistently both the parton saturation effects in photon-proton interaction as well as the nuclear shadowing effects in photon-nucleus process. In comparison to the UPC calculations, we changed the usual photon flux by an effective photon flux, which includes two restrictions: (1) Only photons that hit in the geometrical region of the nucleus-medium are considered and (2) the region of nuclear overlapping is disregarded since we are interested in the coherent photoproduction of the $J/\psi$, which involves the intact part of the nucleus. 

This paper is organized as follows: In the next section we show the main expressions and models used in the rapidity distribution calculation. In the Sec. \ref{eff}, we describe the modification made when the transition from ultraperipheral to peripheral regime occurs. In the Sec. \ref{result}, the main theoretical results are shown. In the last section we summarize the main results and address the conclusions on the study performed.

\section{Theoretical Framework}\label{ipdpf}

\begin{figure}[h]
  \centering
  \scalebox{0.5}{
  \includegraphics{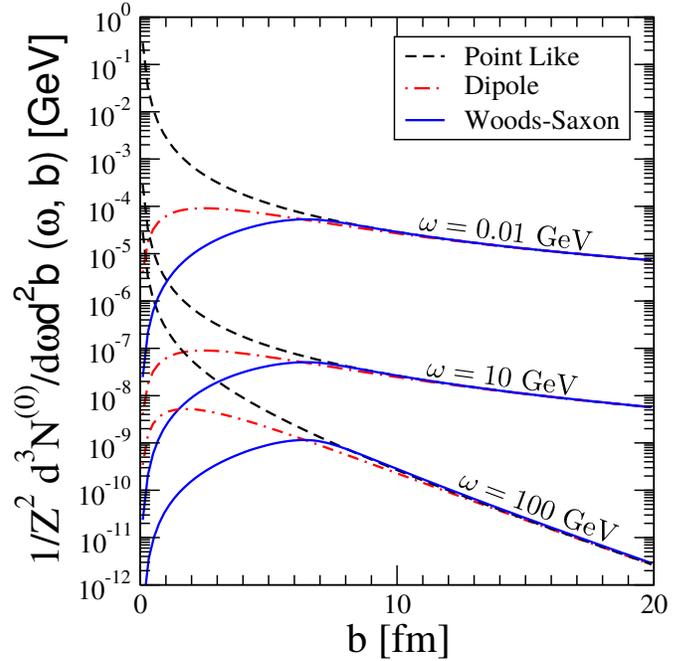}}
  \caption{The b-dependence photon flux distribution for the different form factors of the lead nuclei.\label{fig: figura1}}
\end{figure}

In the ultrarelativistic limit, the rapidity distribution for the vector meson V photoproduction in ultraperipheral collisions AA can be written as a product between an equivalent photon flux, created from one of the nuclei, with the interaction cross section $\gamma A\rightarrow V+A$ \cite{ww}
%\begin{table}[h]
%\centering
%\renewcommand{\arraystretch}{1.2}
%\begin{tabular}{|c|c|c|c|}
%\hline
%\hline
%\multirow{2}{*}{{\color{blue}7 TeV}/{\color{red}13 TeV}} & \multicolumn{2}{|c|}{Boosted-Gaussian} & Gaus-LC \\ \cline{2-4}
%        & $J/\Psi$\, [nb] & $\Psi(2S)$\, [nb] & $J/\Psi$\, [nb] \\ \hline\hline   
%GBW-OLD & \, {\color{blue}5.01}/{\color{red}6.86} \, & \, {\color{blue}0.98}/{\color{red}1.41} \, & \, {\color{blue}4.78}/{\color{red}6.43}\,  \\ \hline
%GBW-KSX & {\color{blue}7.26}/{\color{red}9.66} & {\color{blue}1.70}/{\color{red}2.38} & {\color{blue}6.50}/{\color{red}8.48} \\ \hline
%CGC-OLD & {\color{blue}4.44}/{\color{red}5.80} & {\color{blue}0.87}/{\color{red}1.18} & {\color{blue}4.26}/{\color{red}5.48} \\ \hline
%CGC-NEW & {\color{blue}4.12}/{\color{red}5.43} & {\color{blue}0.85}/{\color{red}1.17} & {\color{blue}3.90}/{\color{red}5.03} \\ \hline 
%BCGC    & {\color{blue}3.79}/{\color{red}4.70} & {\color{blue}0.79}/{\color{red}1.02} & {\color{blue}3.56}/{\color{red}4.36} \\ \hline\hline
%\end{tabular}
%\end{table}
\begin{eqnarray}
\scalebox{0.9}{$\frac{d\sigma}{dy}\left(A+A\rightarrow A+V+A\right)$}&\scalebox{0.9}{$=$}&\scalebox{0.9}{$\omega\frac{dN^{(0)}(\omega)}{d\omega}\sigma_{\left(\gamma A\rightarrow V+A\right)}$}\nonumber
\\
&\scalebox{0.9}{$+$}&\scalebox{0.9}{$\left(y\rightarrow -y\right)$}\label{pri}.
\end{eqnarray}
The factor $dN^{(0)}(\omega)/d\omega$ corresponds to the usual photon flux integrated in the nucleus-nucleus impact parameter b, which depends of the photon energy $\omega$. However, in our calculations, we need a photon flux with b dependence which, according with \cite{PPNP39-503}, can be described using the generic formula
\begin{eqnarray}
\scalebox{0.9}{$\frac{d^3N^{(0)}\left(\omega,b\right)}{d\omega d^2b}=\dfrac{Z^{2}\alpha_{QED}}{\pi^{2}\omega}\left|\int_{0}^{\infty}dk_{\perp}k_{\perp}^{2}\dfrac{F\left(k^2\right)}{k^2}J_{1}\left(bk_{\perp}\right)\right|^{2}$}\label{eq:xx},
\end{eqnarray}
where $Z$ is the atomic number of the nucleus, $F(k^2)$ is the nuclear form factor which represents the nuclear charge distribution and $k^{2}=\left(\frac{\omega}{\gamma}\right)^{2}+k_{\perp}^{2}$, with $\gamma=\sqrt{s_{NN}}/(2m_{\textrm{proton}})$, and $k_{\perp}$ being the transverse momentum of the photon. In the work \cite{PRD94-094023}, it was considered the photoproduction in the ultraperipheral case with $F(k^2)=1$ (point like), resulting in the following photon flux integrated in b, 
\begin{eqnarray}
\scalebox{0.9}{$\frac{dN^{(0)}(\omega)}{d\omega}=\frac{2Z^2\alpha_{em}}{\pi}\left[\chi K_0(\chi)K_1(\chi)-\frac{\chi^2}{2}\left(K_1^2(\chi)-K_0^2(\chi)\right)\right]$},
\end{eqnarray}
where $\chi=2R_A\omega/\gamma$. Now, for the new region of interest, we considered a more realistic dependence of the photon flux, using the form factor obtained from the approximation of the Woods-Saxon distribution as a hard sphere, with radius $R_A$, convoluted with a Yukawa potential with range $a=7$ fm. The Fourier transform of this convolution is the product of the two individual transforms \cite{PRC60-014903}
\begin{eqnarray}
\scalebox{0.9}{$F(k)$}&\scalebox{0.9}{$=$}&\scalebox{0.9}{$\dfrac{4\pi\rho_{0}}{Ak^{3}}\left[\textrm{sin }\left(kR_{A}\right)-kR_{A}\textrm{cos }\left(kR_{A}\right)\right]$}\nonumber
\\
&\scalebox{0.9}{$\times$}&\scalebox{0.9}{$\left[\dfrac{1}{1+a^{2}k^{2}}\right]$},\label{eq:wsy}
\end{eqnarray}
where $A$ is the mass number of the ion and $\rho_{0}=0.1385\textrm{ fm}^{-3}$.
For comparison, we show the dipole form factor often used in the literature and more suited one for small values of $k$ \cite{PRA49-1584}
\begin{eqnarray}
F_{\textrm{dip}}(k^{2})=\frac{\Lambda^2}{\Lambda^2+k^2}, 
\end{eqnarray}
where $\Lambda\approx 88$ MeV for $^{208}Pb$.
In the figure \ref{fig: figura1}, we analyzed the behavior of the photon flux with b dependence for the three form factors presented. It is clear that in the large impact parameter, $b\gtrsim10$ fm, occurs a similar behavior of the photon flux, independent of the form factor used. In contrast, for $b\lesssim 6-7$ fm, the results found by the three models are very different. To understand how these different form factors can affect the two regions of interest (50\%-70\% and 70\%-90\%), it was used the geometrical relation $c=b^2/4R_A^2$ suggested by \cite{PRC65-024905}, which gives an approximated relation between the centrality $c$ and the impact parameter $b$. Applying to our case, the centrality classes 50\%-70\% and 70\%-90\% correspond to $b\simeq 10-11.8$ fm and $b\simeq11.8-13.5$ fm, respectively. Thus, comparing with the Fig. \ref{fig: figura1}, we can see that our results will not be considerably sensitive to the use of these different form factors.  

\begin{figure*}[!ht]
  \centering
  \scalebox{0.5}{
  \includegraphics{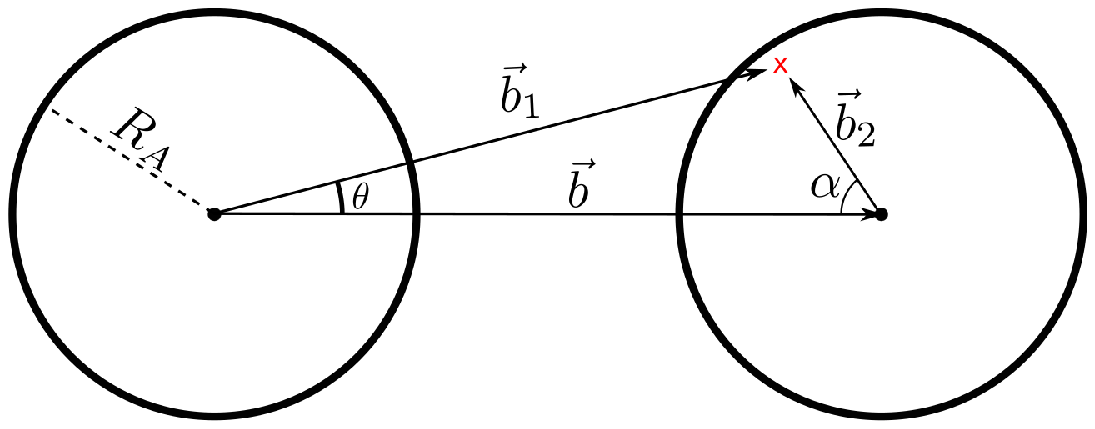}
  \includegraphics{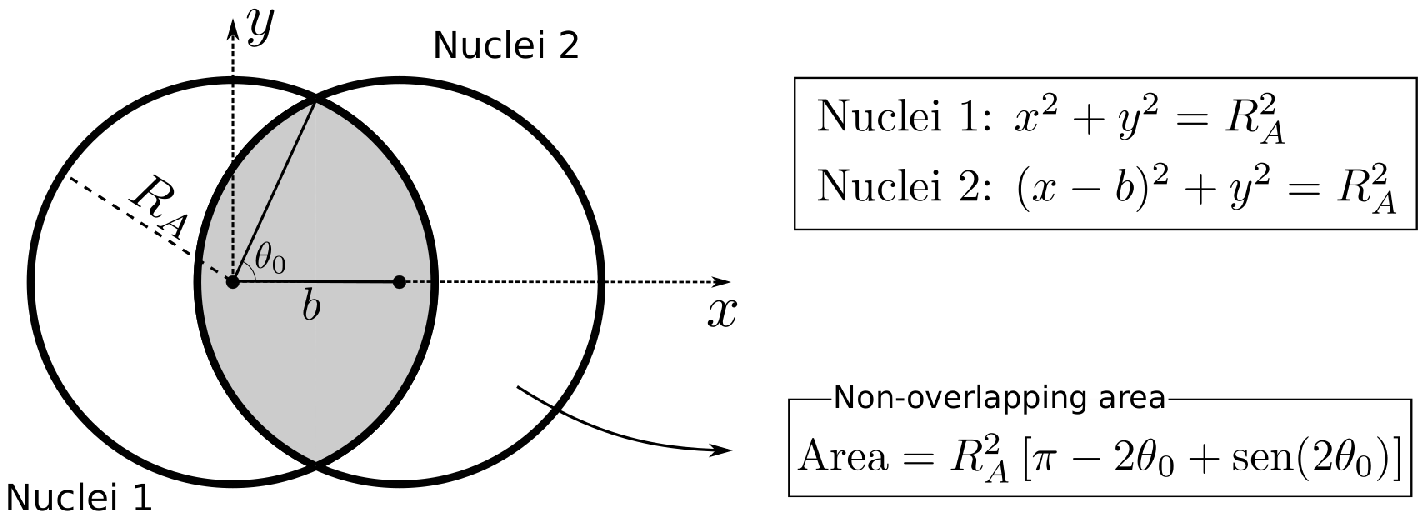}}
  \caption{Left: change of variables $b_{1}\rightarrow b_{2}$ and $\theta\rightarrow\alpha$. Right: sketch of the overlap region existing in the peripheral collisions.}
  \label{newpicture}
\end{figure*}

The second component in the equation (\ref{pri}), $\sigma_{(\gamma A\rightarrow V+A)}$, represents the coherent photonuclear cross section and characterizes the photon-nuclei interaction. In the case which the t-dependence can be factorized, this cross section is defined by 
\begin{eqnarray}
\scalebox{0.9}{$\sigma_{(\gamma A\rightarrow VA)}=\frac{|\textrm{Im }A(x,t=0)|^2}{16\pi}\left(1+\beta^2\right)R^2_g\int_{t_{min}}^{\infty}|F(t)|^2dt$},\label{seg}
\end{eqnarray}
where $\left|\textrm{Im }A(x,t=0)\right|$ represents the imaginary part of the interaction amplitude for the $\gamma A\rightarrow V+A$ process. The parameter  $\beta=\mathrm{Re}A/\mathrm{Im}A$ is necessary to restore the real contribution of the amplitude and usually is defined as \cite{PRD74-074016}
\begin{eqnarray}
\scalebox{0.9}{$\beta = \textrm{tan }\left(\frac{\pi\lambda{eff}}{2}\right)\textrm{ , where }\lambda_{eff}=\dfrac{\partial\textrm{ln }\left[\textrm{Im }A(x,t=0)\right]}{\partial\textrm{ln }s}$}. 
\end{eqnarray}
The second parameter, $R_g^2(\lambda_{eff})$, is important for heavy mesons as $J/\psi$, and corresponds to the ratio of off-forward to forward gluon distribution (skewedness effect), being defined by \cite{PRD60-014015}
\begin{eqnarray}
R_g^2(\lambda_{eff})=\dfrac{2^{2\lambda_{eff}+3}}{\sqrt{\pi}}\dfrac{\Gamma\left(\lambda_{eff}+\frac{5}{2}\right)}{\Gamma\left(\lambda_{eff}+4\right)}. 
\end{eqnarray}
Finally, $F(t)$ is the nuclear form factor integrated from $t_{min}=\left(M_V^2/4\omega\right)^2$.

Based in good results obtained in last works \cite{PRD94-094023,PRD88-017504,PRC89-025201,JPG42-105001}, we described the amplitude $\textrm{Im }A(x,t=0)$ in the colour dipole formalim, where the photon-nuclei scattering can be seen as a sequence of the following subprocesses: (i) the photon fluctuation into quark-antiquark pair (the dipole), (ii) the dipole-target interaction and (iii) the recombination of the $q\bar{q}$ into a vector meson. In this scenary, the amplitude of the process is factorized in the product
\begin{eqnarray}
\scalebox{0.9}{$\textrm{Im }A(x,t=0)=\int d^2r\int\frac{dz}{4\pi}\left(\Psi^*_V\Psi_{\gamma}\right)_T\sigma_{\textrm{dip}}^{\textrm{nucleus}}\left(x,r\right)$},\label{ima}
\end{eqnarray}
where the variables $z$ and $r$ are the longitudinal momentum fraction carried by the quark and the transverse color dipole size, respectively.

The transverse overlap of the photon-meson wave function, $\left(\Psi^*_V\Psi_{\gamma}\right)_{T}$, can be written as \cite{PRD74-074016}
\begin{eqnarray}
\scalebox{0.9}{$\left(\Psi^*_V\Psi_{\gamma}\right)_T$} &\scalebox{0.9}{$=$}& \scalebox{0.9}{$\hat{e}_fe\frac{N_c}{\pi z(1-z)}\left\{m^2_fK_0(\epsilon r)\phi_T(r,z)\right.$}\nonumber
\\
&\scalebox{0.9}{$-$}&\scalebox{0.9}{$\left.\left[z^2+(1-z)^2\right]\epsilon K_1(\epsilon r)\partial_r\phi_T(r,z)\right\}$}
\end{eqnarray}
where the phenomenological term $\phi_T(r,z)$ represents the scalar part of the meson wave-function. Here, it was used the Boosted-Gaussian model \cite{ZPC75-71} since it can be applied in a systematic way for the excited states. The parameters $\mathcal{R}^2_{nS}$ and $\mathcal{N}_{nS}$ presented in the model can be found in \cite{PRD90-054003,Sanda2}.

The next term in the equation (\ref{ima}) is the cross section $\sigma_{\textrm{dip}}^{\textrm{nucleus}}\left(x,r\right)$, calculated via Glauber model \cite{EPJC26-35}, 
\begin{eqnarray}
\scalebox{0.9}{$\sigma_{\textrm{dip}}^{\textrm{nucleus}}\left(x,r\right)$}&\scalebox{0.9}{$=$}&\scalebox{0.9}{$2\int d^2b$}\nonumber
\\
&\scalebox{0.9}{$\times$}&\scalebox{0.9}{$\left\{1-\textrm{exp}\left[-\frac{1}{2}T_A(b)\sigma_{\textrm{dip}}^{\textrm{proton}}\left(x,r\right)\right]\right\}$}
\end{eqnarray}
where the nuclear profile function, $T_A(b)$, will be obtained from a 3-parameter Fermi distribution for the nuclear density \cite{Fermi}. The dipole cross section, $\sigma_{\textrm{dip}}^{\textrm{proton}}(x,r)$, is related to the dipole-proton scattering amplitude in the form $\sigma_{q\bar q}(x,r) = 2\,\int d^2b \,A_{q\bar{q}}(x,r,b)$, bearing in mind that $b$ and $\Delta$ are Fourier conjugate variables. There are different models for the amplitude $A_{q\bar{q}}(x,r,b)$, and here, it was considered the model GBW \cite{PRD59-014017} since that in previous works (ex. \cite{PRD94-094023}) we did not see great variation between models like GBW, IIM \cite{PLB590-199} and IIM with b dependence \cite{PRD74-074016}, for the rapidity distribution. 

\section{The Effective Photon Flux}\label{eff}

Following the reference \cite{PRC93-044912}, the effective photon flux can be constructed from the usual photon flux as
\begin{equation}
\scalebox{0.9}{$N^{(2)}\left(\omega_{1},b\right)=\int N\left(\omega_{1},b_{1}\right)\frac{\theta(R_{A}-b_{2})\times\theta(b_{1}-R_{A})}{A_{eff}(b)}d^{2}b_{1}$}\label{eq:fluxoefetivo}
\end{equation}
where we modify the original equation by applying the effective area, $A_{eff}(b)$, in contrast to the fixed value $\pi R_{A}^2$ present in \cite{PRC93-044912}. The function $\theta\left(R_{A}-b_{2}\right)$ ensures that the effective photon flux will only be formed by photons that reach the geometrical region of the target-nuclei, while the function $\theta(b_{1}-R_{A})$ disregards the overlap region where the nuclear effects are present. To eliminate the step functions, it was performed the variables substitution $b_{1}\rightarrow b_{2}$ and $\theta\rightarrow\alpha$, represented in the Fig. \ref{newpicture}.
%\begin{figure}[h]
%  \centering
%  \scalebox{0.4}{
%  \includegraphics{N2nucleo.eps}}
%  \caption{Change of variables: $b_{1}\rightarrow b_{2}$ and $\theta\rightarrow\alpha$.}
%  \label{newpicture}
%\end{figure}

\begin{figure}[h]
  \centering
  \scalebox{0.4}{
  \includegraphics{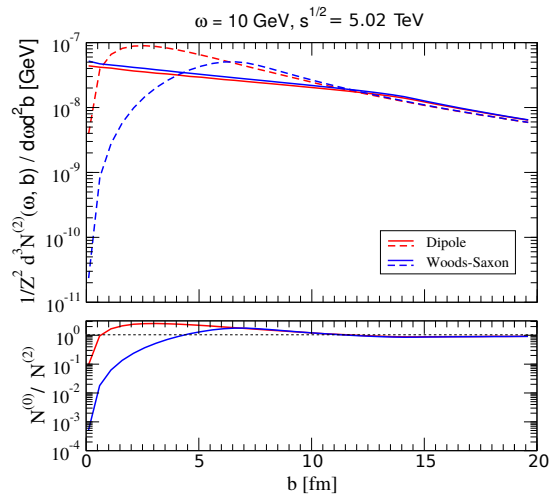}}
  \caption{Comparison between the usual photon flux (dashed line) and the effective photon flux (solid line) for the photon energy $\omega=10$ GeV. \label{fig: figura1-1}}
\end{figure}

In terms of the new variable, the equation (\ref{eq:fluxoefetivo}) can be rewritten as 
\begin{eqnarray}
\scalebox{0.9}{$N^{(2)}\left(\omega_{1},b\right)$}&\scalebox{0.9}{$=$}&\scalebox{0.9}{$\frac{1}{A_{eff}(b)}\left[\int_{0}^{2\pi}\int_{0}^{R_{A}}N\left(\omega_{1},b_{1}\right)b_{2}db_{2}d\alpha\right.$}\nonumber
\\
&\scalebox{0.9}{$-$}&\scalebox{0.9}{$2\int_{0}^{\sqrt{R_{A}^{2}-b^{2}/4}}db_{1y}$}\nonumber
\\
&\scalebox{0.9}{$\times$}&\scalebox{0.9}{$\left.\int_{-\sqrt{R_{A}^{2}-y^{2}}+b}^{\sqrt{R_{A}^{2}-y^{2}}}db_{1x}N\left(\omega_{1},b_{1}\right)\right]$}\label{eq:fluxonovo}
\end{eqnarray}

where \scalebox{0.7}{$A_{eff}(b)=R_A^2\left[\pi-2\textrm{arccos}\left(\frac{b}{2R_A}\right)\right]+\frac{b}{2}\sqrt{4R_A^2-b^2}$} is the considered effective area, $b_1=\sqrt{b_2^2+b^2-2b_2b\textrm{cos }(\alpha)}$ in the first term and $b_1=\sqrt{b_{1x}^2+b_{1y}^2}$ in the second term. In (\ref{eq:fluxonovo}), the first term acts only on the geometrical region of the target-nuclei, while the second term disregards the overlap region of the nucleus. 

Using the equation (\ref{eq:fluxonovo}), it was obtained the Fig. \ref{fig: figura1-1}, where we compare the effective photon flux with the usual one for the photon energy $\omega=10$ GeV, which corresponds to a meson rapidity $y=\textrm{ln}\left(2\omega/m_V\right)\simeq 1.85$. In the first region (50\%-70\%), the usual photon flux is slightly larger than the effective photon flux. The opposite occurs in the second region (70\%-90\%), where the overlap term is small, tending to unity as we move towards the ultraperipheral region. Thus, one should not expect a large variation in the transition from the usual photon flux to the effective photon flux in the analyzed region 50\%-90\%. 

\section{Results and discussions}\label{result}

\begin{figure}[h]
\centering
\scalebox{0.8}{
\includegraphics{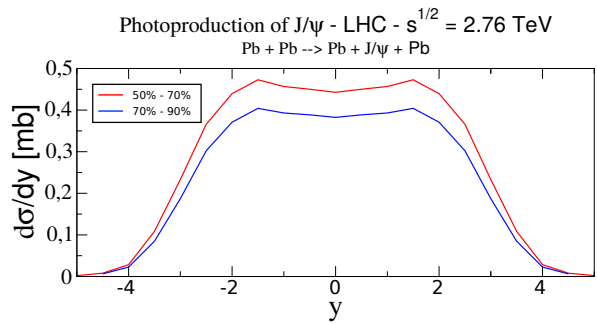}}
\caption{Rapidity distribution for $J/\Psi$ photoproduction at $\sqrt{s}=2.76$ TeV using the GBW dipole model.}
\label{psi1s}
\end{figure}

In this section, we present the results of the rapidity distribution for the photoproduction in Pb-Pb collisions of $J/\Psi$ states in the centrality regions 50\%-70\% and 70\%-90\%, at the energy $\sqrt{s}=2.76$ TeV and $\sqrt{s}=5.02$ TeV. In our calculations, it was applyed the form factor (\ref{eq:wsy}), which is more apropriated for the heavy nucleus, although, as pointed out in the section (\ref{ipdpf}), it is not expected a considerable change. Firstly, in the Fig. \ref{psi1s}, are presented the results for $J/\Psi$ at $\sqrt{s}=2.76$ TeV for the two analyzed regions using the GBW dipole model. The results between the different centrality classes are similar in behavior, with maximum difference varying from 15\% in $y=0$ to 27\% in rapidity $|y|\approx 3.5$. The difference in the results allows the future comparison with data, given us information about how far the used formalism can be extrapolated and trusted.

\begin{figure}[h]
\centering
\scalebox{0.8}{
\includegraphics{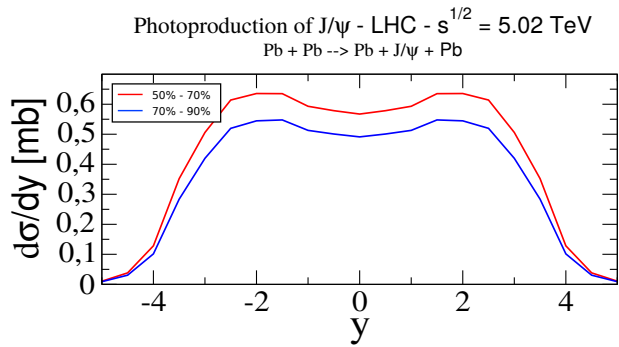}}
\caption{Rapidity distribution for $J/\Psi$ photoproduction at $\sqrt{s}=5.02$ TeV using the GBW dipole model.}
\label{psi1s5020}
\end{figure}

In our second pair of results presented in Fig. \ref{psi1s5020}, it was calculated the $J/\Psi$ production at energy $\sqrt{s}=5.02$ TeV. As in the previous case, the difference in the results between the two centrality classes varies from 15\% in $y=0$ to 26\% in rapidity $|y|\approx 4$. 

We also calculated the ratio $\frac{d\sigma^{5.02}}{dy}/\frac{d\sigma^{2.76}}{dy}$ and obtained an increase of approximately 30\% in the central rapidity region $|y|<1.5$ for the two centrality classes analyzed. This same ratio is approximately 60\% for the same rapidity region in UPC. Thus, the effective photon flux appears to be less sensitive to the variation of energy in relation to usual photon flux. On the other hand, in the model adopted here for the transition from the ultraperipheral to the peripheral regime, no modification was made in the photonuclear cross section since the variation in the nucleus-nucleus impact parameter affects mainly the photon flux. The photonuclear cross section is calculated using the Glauber model which, in turn, is related to the number of nucleons that interact with the photon, then a certain modification could be expected in the peripheral case where the number of nucleons is smaller.

\section{Summary}

We have considered the coherent photoproduction of $J/\psi$ state in peripheral $Pb-Pb$ collision at LHC using the color dipole approach as the underlying theoretical framework. The rapidity distributions in the centrality classes 50\%-70\% and 70\%-90\% have been presented, allowing us to test the robustness of the dipole formalism. In our peripheral calculations, we consider a modified photon flux without change of the photonuclear cross section in relation to the ultraperipheral (UPC) case. From this approach, it was verified that in the region analyzed the application of the effective photon flux does not result in a considerable change in the results in relation to the usual photon flux. Othewise, a more dramatic change will occur in a more central region. However, it could deserve a more sophisticated undertanding of the behavior of the photonuclear cross section in peripheral collisions, in order to carry out a more complete and reliable analysis. The point here was to start the study about the contribution of the photoproduction in more central collisions, which is an analysis still not much explored in the literature. The constraints of this calculation require the onset of new data.

\begin{acknowledgments}
We would like to thank Dr.\hspace{1mm}Ionut Arsene for usefull discussions. This work was partially financed by the Brazilian funding agency CNPq.

\end{acknowledgments}

\end{document}